\documentclass[aps,pre,onecolumn,showpacs,amsmath,amssymb,longbibliography,notitlepage,11pt,tightenlines,superscriptaddress]{revtex4-1} 
\usepackage{xcolor}
\usepackage{graphicx}
\usepackage{setspace}
\usepackage{changepage}

\definecolor{darkblue}{rgb}{0,0,0.6}
\definecolor{darkred}{rgb}{0.6,0,0}
\usepackage[colorlinks=true,urlcolor=darkblue,citecolor=darkblue,linkco
lor=darkred,hyperfootnotes=false]{hyperref}

\newcommand{\ind}[1]{_{\mathrm{#1}}}
\DeclareMathOperator{\Ai}{Ai}
\newcommand{\lcurv}{\ell_\mathrm{curv}}
\newcommand{\Rf}{R}
\newcommand{\srr}{\sigma_{rr}}
\newcommand{\stt}{\sigma_{\theta\theta}}

\newcommand{\SIsection}[1]{\vspace{10pt}\noindent\textsf{\textbf{\normalsize#1}}\vspace{2pt}}
\newcommand{\SIsubsection}[1]{\vspace{5pt}\noindent\textsf{\textbf{#1.}}}

\begin{document}

\title{Geometry underlies the mechanical stiffening and softening of an indented floating film\footnote{Full text available at: https://doi.org/10.1039/D0SM00250J}}
\author{Monica M. Ripp}
\affiliation{Department of Physics, Syracuse University, Syracuse, NY 13244, USA}
\affiliation{BioInspired Syracuse: Institute for Material and Living Systems, Syracuse University, Syracuse, NY 13244, USA}
\author{Vincent D\'emery}
\email{vincent.demery@espci.psl.eu}
\affiliation{Gulliver, CNRS, ESPCI Paris, PSL Research University, 10 rue Vauquelin, 75005 Paris, France}
\affiliation{Univ Lyon, ENS de Lyon, Univ Claude Bernard Lyon 1, CNRS, Laboratoire de Physique, F-69342 Lyon, France}
\author{Teng Zhang}
\email{tzhang48@syr.edu}
\affiliation{BioInspired Syracuse: Institute for Material and Living Systems, Syracuse University, Syracuse, NY 13244, USA}
\affiliation{Department of Mechanical and Aerospace Engineering, Syracuse University, Syracuse, NY 13244, USA}
\author{Joseph D. Paulsen}
\email{jdpaulse@syr.edu}
\affiliation{Department of Physics, Syracuse University, Syracuse, NY 13244, USA}
\affiliation{BioInspired Syracuse: Institute for Material and Living Systems, Syracuse University, Syracuse, NY 13244, USA}

\begin{abstract}
A basic paradigm underlying the Hookean mechanics of amorphous, isotropic solids is that small deformations are proportional to the magnitude of external forces. However, slender bodies may undergo large deformations even under minute forces, leading to nonlinear responses rooted in purely geometric effects. Here we study the indentation of a polymer film on a liquid bath. Our experiments and simulations support a recently-predicted stiffening response [Vella \& Davidovitch, \textit{Phys.~Rev.~E} \textbf{98}, 013003 (2018)], and we show that the system softens at large slopes, in agreement with our theory that addresses small and large deflections. We show how stiffening and softening emanate from nontrivial yet generic features of the stress and displacement fields. 
\end{abstract}

\maketitle

\section{Introduction}

\noindent A major challenge in the mechanics of materials and structures is bridging the gap between a system's local material response and its global stiffness. 
This connection from microscopic to macroscopic scales is often complicated by subtle geometric effects \cite{Paulsen19}. 
One conceptually simple example is a buckled elastic rod, which drastically changes its shape in response to loading. 
Such \emph{elastica} problems captured the attention of Galileo, the Bernoullis, and Euler, and variations on them continue to fascinate and push our understanding of slender bodies today \cite{Flaherty72,Audoly10,Giomi12,Davidovitch19}. 
Particular attention surrounds the behaviors of two-dimensional sheets, which may carry tensile loads in one direction while buckling or wrinkling in a perpendicular direction, leading to large anisotropies in the equilibrium stresses and deformations 
\cite{Davidovitch11,Li12,Hure12,Bella14,Chopin19}. 
Understanding such geometrically-nonlinear behaviors is important to a wide range of applications, from stretchable electronics \cite{Rogers10} to large-scale inflatable structures \cite{Pagitz07}.

\begin{figure}[t]
\centering
\includegraphics[width=0.6\linewidth]{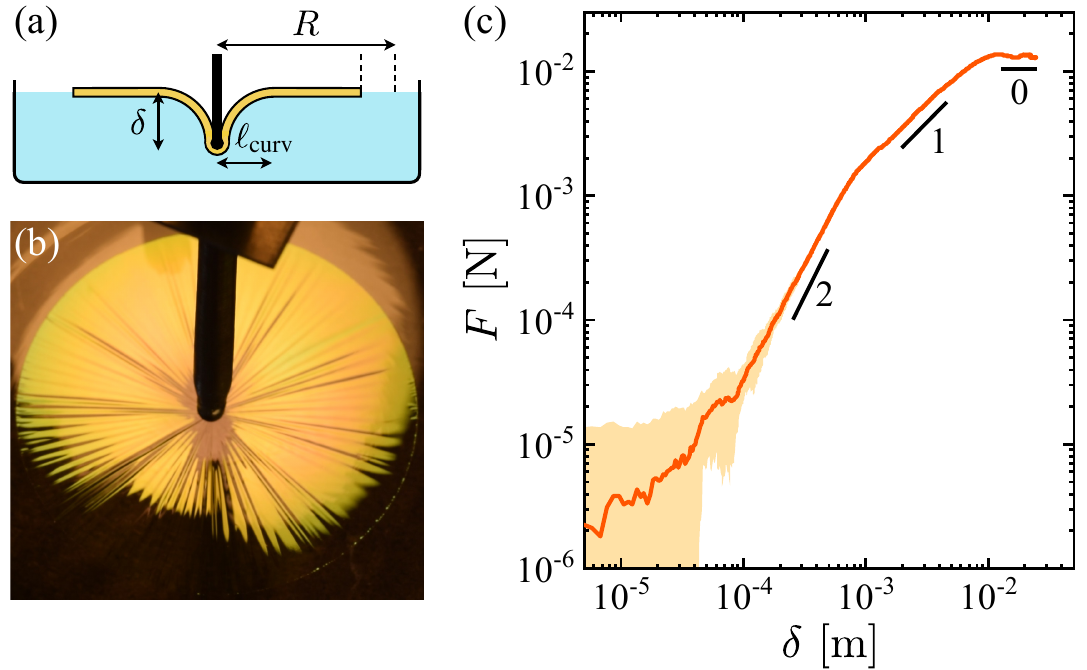}
\caption{
\textbf{Indentation of a floating circular polymer film of radius $R$.} 
\textbf{(a)} We measure the normal force, $F$, versus indentation depth, $\delta$. 
\textbf{(b)} A film of thickness $t=445$ nm and radius $\Rf=22$ mm indented to depth $\delta=0.9$ mm, causing radial wrinkles to cover the sheet. 
\textbf{(c)} Measured $F(\delta)$ for a film with $t=213$ nm and $\Rf=44$ mm, where the shaded band shows experimental uncertainty due to noise. 
The data show a complex response with multiple distinct scaling regimes. 
At large $\delta$, the system energy is dominated by the cost of exposing additional liquid-air interfacial area as the boundary of the film is pulled inwards, illustrated by the vertical dashed lines in (a). 
This geometric effect leads to a plateau in the force (\textit{i.e.,} softening). 
}
\label{fig:1}
\end{figure}

Here we study the indentation of a polymer film on a liquid bath [Fig.~\ref{fig:1}a,b] using experiments and simulations spanning four decades in indentation depth, and theory that addresses small and large deflections. 
This model system shows a remarkably rich response under continual loading---it first stiffens (\textit{i.e.}, $F/\delta$ increases) and then subsequently softens [Fig.~\ref{fig:1}c]. 
Whereas stiffening was predicted by a recent detailed theoretical study \cite{Vella18}, softening occurs at large slopes where a direct analysis of the F\"oppl--von K\'arm\'an equations is prohibitive. 
We harness a simple geometric model that treats the sheet as inextensible but with zero bending cost \cite{Paulsen15}, which allows us to understand both the force and the film profile at large amplitude. 

Particular interest in the indentation of thin films surrounds the observation that wrinkles significantly modify the stress field in the sheet and the force transmitted to the indenter \cite{Vella15,Vella17,Vella18}, which has practical importance for indentation assays used to measure film properties. 
Fundamental interest is bolstered by the recent discovery of a novel wrinkled morphology that is nearly isometric to the original undeformed state \cite{Vella15,Vella15b}. 
Such ``asymptotically isometric mechanics'' arises in the dual limit of weak applied tension and vanishing bending modulus. 
Experiments on ultrathin polymer films measuring the extent of wrinkles \cite{Vella15}, the vertical profile of the film \cite{Paulsen16}, and the force response \cite{Holmes10} support this picture. 
Yet, apart from this particular regime, direct measurements of forces in wrinkled interfacial films are scarce. 
Here we confirm the entire predicted force evolution for an indented floating film \cite{Vella15,Vella18}, and we demonstrate an additional regime at large amplitude with two surprising features: A force plateau and a limiting interface shape that is independent of the indentation depth. 
Our results establish the maximal load-bearing capacity of a floating film under point loading, and our theoretical approach for small and large deflections can readily be adapted to different geometries and loading conditions. 
We close by highlighting analogous behaviors that arise from similar mechanisms in model fiber networks, due to a separation of energy scales between bending and stretching deformations that favors nearly-isometric wrinkled deformations in a sheet and the buckling of individual fibers in a network.

\section{Model system}

\noindent We work in a geometry previously investigated by Holmes and Crosby \cite{Holmes10} and Huang \cite{Huang10}. 
We use spin-coated, ultrathin polystyrene sheets (thickness $58 < t < 490$ nm, Young's modulus $E = 3.4$ GPa) of circular shape (radius $11 < \Rf < 44$ mm), that are floated on a liquid bath of density $\rho$ = 1000 kg/m$^3$ and surface tension $\gamma = 72$ mN/m. 
We indent the films a vertical distance $\delta$ as illustrated in Fig.~\ref{fig:1}a, using a custom force probe mounted on a computer-controlled vertical translation stage, following Ref.~\citealp{Holmes10}. 
To access an even larger dynamic range of indentation depths, we perform numerical simulations in the molecular dynamics software LAMMPS \cite{Plimpton95}. 
We use a triangular lattice model for the sheet \cite{Seung88} and springs with zero rest length for the liquid \cite{Giomi12}. 
This approach allows us to use parallel computing methods to address this inherently multiscale problem where the sheet thickness, wrinkle wavelength, and sheet radius occupy separate lengthscales. 

Our films fall in the doubly-asymptotic limit of weak tension, $\gamma/Y < 10^{-3}$, and negligible bending stiffness, $\epsilon = B \rho g/\gamma^2 < 10^{-4}$, where $Y = Et$ is the stretching modulus, $B$ is the bending modulus, $g$ is the gravitational acceleration, and $\epsilon$ is the inverse bendability \cite{Davidovitch11}. 
Such films can bare only a minute level of compression before buckling out of plane, and they exhibit an approximately linear stress-strain response under tensile loading, all the way up to fracture \cite{Liu15}. 
We work in the regime $\Rf \gg \ell_c$, where $\ell_c =\sqrt{\gamma/\rho g} \approx 2.7$ mm is the gravity-capillary length, which defines the lateral scale on which a bare liquid interface would be disturbed due to a point deflection. 

As the sheet is indented beyond a threshold $\delta$, wrinkles form in an annular region and then expand radially towards the indenter and the edge of the film~\cite{Vella15,Paulsen16}, eventually covering the film as in Fig.~\ref{fig:1}b. 
Figure~\ref{fig:1}c shows the measured force versus displacement for an experiment with $t=213$ nm and $\Rf=44$ mm. 
The data show a complex response with multiple distinct scaling regimes. 
At the largest $\delta$ probed, the force levels off to a constant value. 
For a bulk material, such a plateau would suggest plasticity or failure, but as we will show, here the coupling of the geometry of the sheet and the transmission of forces allows such a response within the Hookean regime of the material. 
At smaller $\delta$ the force $F(\delta)$ is linear in $\delta$, but this is not the conventional linear response often obtained at infinitesimal amplitude. 
At yet smaller $\delta$ the force is markedly nonlinear and proportional to $\delta^2$. 
For $\delta \lesssim 10^{-4}$~m, our experimental uncertainties are too large to extract a clear scaling; 
here our simulations are able to resolve a transition to an additional linear regime for very small $\delta$ 
[Fig.~\ref{fig:4}b]. 
We thus observe four distinct scaling regimes for the normal force. 

This phenomenology is consistent with a recent detailed theoretical analysis of indentation~\cite{Vella18}, although that work was limited to small slopes and did not anticipate the force plateau at large $\delta$. 
We begin by reviewing key results from previous theory and then develop a complementary theoretical approach that is valid for arbitrary slopes and explains the observed behavior at large amplitude.

\section{Theory}

\subsection{Previous results}
Progress on understanding the mechanical response of thin floating films to indentation has relied on a far-from-threshold (FT) framework that was developed in a series of recent works~\cite{Davidovitch11,Davidovitch12,Schroll13,Chopin15}. 
This theory is based on the observation that for sufficiently thin films, out-of-plane buckling completely relaxes the compressive stress in one direction. 
In this framework, complex wrinkle patterns emerge from at least two distinct competitions: (i) in-plane stretching energy determines the extent of the wrinkled zone; 
(ii) inside this region, the wrinkle wavelength comes from balancing bending of the sheet with an effective substrate stiffness~\cite{Cerda03,Paulsen16}. 
(Studies of ordering on other mesoscopic lengthscales are ongoing~\cite{Tovkach20}.) 

In the present problem, the indenter performs work that is transmitted to a combination of elastic stretching of the sheet, gravitational energy of the displaced fluid, and surface energy due to exposing liquid surface area as the sheet retracts radially inward [dashed lines in Fig.~\ref{fig:1}a]. 
At small $\delta$, the indenter probes the stress state of the film, which is set by the interfacial tension pulling at its edge. 
Thus, in the incipient regime: 
\begin{equation}
F \simeq \frac{4\pi \gamma}{\ln (1/\epsilon)} \delta \ \ \ \text{(regime I)}. 
\label{linear}
\end{equation}
Here the film is stretched in a central region of width comparable to the gravity-capillary length, $\ell_c = \sqrt{\gamma/\rho g}$~\footnote[3]{There are logarithmic corrections to Eq.~(\ref{linear}) for a finite-size indenter; see for example ref~\cite{Vella17}.}. 

At larger $\delta$ the deflections in the sheet become nonlinear and the force departs from this scaling. 
Eventually, azimuthal stresses are compressive in an annular region, leading to the formation of wrinkles that \textit{qualitatively} modify the global stress field. 
When wrinkles cover a finite fraction of the sheet, the force is predicted to be: 
\begin{equation}
F \simeq 2.26 \,\sqrt{ \frac{ Y \rho g }{ \ln( \tilde{\delta} ) } } \delta^2 \ \ \ \text{(regime II)},
\label{quadratic}
\end{equation}
where $\tilde{\delta} = (\delta/\ell_c) \sqrt{Y/\gamma}$ is a dimensionless indentation depth that compares geometric and mechanical strain~\cite{Vella15}. 
Note that Eq.~(\ref{quadratic}) is a stiffening response because the effective spring constant $F/\delta$ (hereafter called the \textit{stiffness}) is an increasing function of $\delta$. 

At still larger $\delta$, wrinkles reach the edge of the sheet. 
This event gives rise to a so-called ``asymptotic isometry" where elastic energies in the sheet become negligible. 
In this regime, the vertical profile of the sheet decays over an emergent lateral scale $\ell_\text{curv} = \ell_c^{2/3} \Rf^{1/3}$ to minimize the sum of gravitational and surface energies~\cite{Vella15,Paulsen16}. 
A precise calculation yields: 
\begin{equation}
F \simeq 4.58 \,(\gamma \Rf)^{2/3} (\rho g)^{1/3} \delta \ \ \ \text{(regime III)} . 
\label{quasilinear}
\end{equation}
We note that regime I can be understood with standard linear response theory, whereas regime III arises from distinct geometric effects~\cite{Vella15,Vella18}, yet the force is proportional to $\delta$ in both regimes. 

The transitions between these regimes are governed by the evolution of the stresses in the film~\cite{Vella18}. 
A broad transition to regime II is predicted at indentation depth: 
\begin{equation}
\delta_{*} \sim \frac{\gamma}{\sqrt{Y \rho g}} \ \ \ \text{(I $\rightarrow$ II)} , 
\label{delta*}
\end{equation}

\noindent with a numerical prefactor reported in Ref.~\citealp{Vella18} that varies with $\epsilon^{-1}$. 
Regime III is then predicted to begin at: 
\begin{equation}
\delta_{**} \simeq 2.63 \,\Rf^{2/3} \ell_c^{1/3} \sqrt{ \frac{ \gamma }{ Y } \ln \left( \frac{\Rf}{\ell_c} \right) } \ \ \ \text{(II $\rightarrow$ III)} . 
\label{delta**}
\end{equation}

With the exception of Eq.~(\ref{quasilinear}), the above predictions have not been tested by experiments or simulations~\footnote[4]{Equation~(\ref{quasilinear}) was compared with the experiments of Holmes \& Crosby~\cite{Holmes10} in the SI to Ref.~\citealp{Vella15}.}; we do so in Sec.~\ref{sec:comparison}. 
Moreover, the behaviors at large slopes are completely unexplored. 
As we will show, the force plateau at large amplitude constitutes another distinct scaling regime, which we define as regime IV.

\subsection{Stress field and indentation force}
\noindent Here we outline a set of general arguments based on force balance that are applicable to indentation at both small and large slopes. 
This treatment allows us to describe the fundamental mechanisms for the previously-predicted stiffening at small slopes~\cite{Vella18}, while elucidating a novel softening regime at large slopes. 

We consider axially-symmetric vertical deflections of the sheet. 
Radial force balance in the sheet reads: $\frac{d}{dr}(r\srr) -\stt = 0$, where $\srr$ and $\stt$ are the radial and azimuthal stress components, with the boundary condition $\srr(\Rf)=\gamma$ at the edge. 
This equation has two simple solutions: 
In the undeformed state, the stress field is uniform, corresponding to the first solution: $\srr=\stt=\gamma$. 
At large indentation (regimes III and IV), radial wrinkles cover the entire sheet to avoid azimuthal compression that would have otherwise been induced by the contraction of circles, so that $\stt = 0$ \cite{Vella18}. 
Thus, in the second solution: 
\begin{equation}\label{eq:srr_III}
\srr(r) = \frac{\gamma\Rf}{r}. 
\end{equation}
Here, the sheet behaves as if it were composed of many radial strings transmitting the stress from the boundary to the indenter. 

\begin{figure}[tb]
\centering
\includegraphics[width=6.5cm]{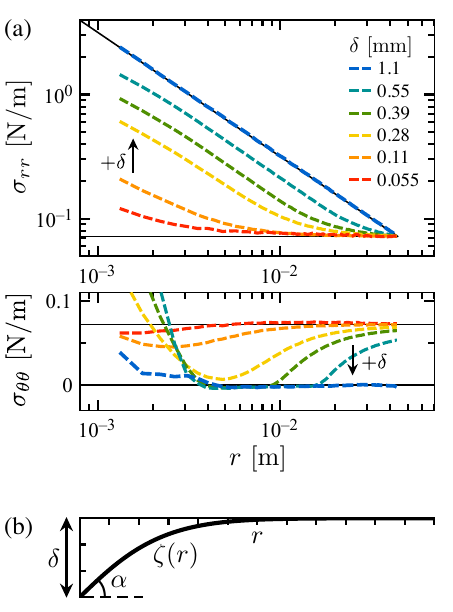}
\caption{
\textbf{Radial stress and vertical profile.} 
\textbf{(a)} Radial tensile stress, $\srr(r)$, and hoop stress, $\stt$, measured in simulations with $t=210$ nm, $\gamma=72$ mN/m, $\Rf=44$ mm, and averaged over $\theta$. 
As $\delta$ increases, vertical deflections of the film reduce hoop stresses relative to radial stresses, amplifying the radial stress at the indenter. 
Solid lines show the limiting behaviors at $\delta=0$ [where $\srr(r) = \stt(r) = \gamma$] and at large indentation [$\srr(r) = \gamma\Rf/r$ and $\stt(r) = 0$]. 
\textbf{(b)} Vertical profile, $\zeta(r)$, shown in the shape of an Airy function. 
The displacement at the origin is $\delta$, where the sheet forms an angle $\alpha$ with the horizontal. 
We take the profile of the gross shape to be axisymmetric; when this symmetry is broken there will be corrections to the force relations derived here. 
}
\label{fig:2}
\end{figure}

Figure~\ref{fig:2}a shows the radial tensile stress $\srr$ measured in our simulations. 
As $\delta$ increases, the data go away from the uniform solution towards the second solution that transmits larger stresses to the center of the sheet. 
This occurs as the hoop stress $\stt$ drops from $\gamma$ towards $0$ over the majority of the film [Fig.~\ref{fig:2}a, lower panel]. 

The radial tensile stress is linked to the normal force on the indenter, $F$, by vertical force balance: $F = 2\pi \lim_{r\to 0} (r\srr)\sin(\alpha)$, where $\alpha$ is the angle between the horizontal and the sheet at $r=0$, as drawn in Fig.~\ref{fig:2}b. 
We denote the axisymmetric height of the sheet by $\zeta(r)$, which neglects the undulations due to small-amplitude wrinkles, so $\sin(\alpha)=\zeta'(0)/\sqrt{1+\zeta'(0)^2}$. 
In regimes III and IV we may use Eq.~(\ref{eq:srr_III}) for the radial stress, leading to: 
\begin{equation}
F = 2\pi\Rf \gamma\sin(\alpha). 
\label{force-angle}
\end{equation}
These basic considerations get at the essence of the observed stiffening and softening responses. 
Reduction of the hoop stress due to radial retraction leads to larger radial stresses; this growing anisotropy of stresses causes stiffening. 
Then at large slopes, the force saturates as $\sin(\alpha)\simeq 1$ [via Eq.~(\ref{force-angle})], which causes softening. 
To substantiate this qualitative picture, we now consider the response of the film in more detail, starting from small indentation.

\subsection{Stiffening due to growing anisotropy of stresses}
We begin by estimating the force on the indenter within the linear response of the system at infinitesimal indentation. 
Here the stress in the film is given by the simple isotropic solution, namely, $\srr=\stt=\gamma$. 
We suppose that the indenter will deform a finite region of the sheet, so that the stress and displacement are modified only within a core of some finite radius $\ell$ (where it is natural to assume that $\ell \propto \ell_c$ since there is no other available length). 
The force transmitted to the indenter is then estimated by $2\pi \ell \gamma \sin(\alpha)$, in analogy with Eq.~(\ref{force-angle}) but with $\sin(\alpha) \simeq \delta/\ell$. 
Thus, in the initial regime, the force on the indenter should scale as $F \sim \gamma \delta$, independent of $\ell$ so long as $\ell < \Rf$. 
This scaling is in agreement with Eq.~(\ref{linear}). 

Intuitively, moving from an approximately uniform stress field~\footnote[5]{In the approximately uniform solution, there is nevertheless a stretched core region where the stress is transmitted to the indenter~\cite{Vella18}.} to an increasingly anisotropic stress field causes the observed stiffening, as forces are transmitted more effectively to the indenter in the latter case. 
This qualitative picture is supported by a recent far-from-threshold analysis of the F\"oppl--von K\'arm\'an equations \cite{Vella15,Vella18}, summarized above. 
At larger $\delta$, wrinkles reach the edge of the sheet, giving rise to a so-called ``asymptotic isometry'' \cite{Vella15,Vella18} where elastic energies in the sheet become negligible.

\subsection{Softening due to large slopes}
We now show how large slopes give rise to a previously-unanticipated softening response. 
We begin by considering the case where wrinkles cover the sheet, yet the vertical deflections are still within the small-slopes limit. 
Under these conditions, the vertical profile is given by: 
\begin{equation}\label{Airy}
\zeta(r) = -\delta \frac{\Ai(r/\lcurv)}{\Ai(0)},
\end{equation}
where $\Ai(r)$ is the Airy function shown by the curve in Fig.~\ref{fig:2}b, and $\lcurv = \ell_c^{2/3}\Rf^{1/3}$. 
This profile may be derived by solving for the stresses and deflections everywhere in the sheet \cite{Vella15} or by minimizing the surface and gravitational energy using the geometric model described in Sec.~\ref{sec:geom}. 
Using this result, the slope at the origin is $\zeta'(0) = -(\delta/\lcurv) \Ai'(0)/\Ai(0) \simeq 0.729 (\delta/\lcurv)$, so that small slopes are obtained when $\delta\ll\lcurv$. 
Equation~(\ref{force-angle}) for the force then reduces to: 
\begin{equation}
F = 2\pi\Rf\gamma \zeta'(0) \simeq 4.58 \, (\gamma \Rf)^{2/3} (\rho g)^{1/3} \delta \ \ \ \text{(regime III)} . 
\end{equation}
This result matches Eq.~(\ref{quasilinear}), which was predicted in Ref.~\citealp{Vella15} by different means. 

Our approach can also address large slopes, which correspond to $\delta\gtrsim\lcurv$. 
The asymptotic force is obtained by setting $\sin(\alpha) = 1$ in Eq.~(\ref{force-angle}), yielding:
\begin{equation}\label{constant}
F=2\pi\Rf\gamma \ \ \ \text{(regime IV)}. 
\end{equation}
This is a softening response because the stiffness, $F/\delta$, is a decreasing function of $\delta$. 
The crossover occurs when the expressions (\ref{quasilinear}) and (\ref{constant}) are comparable:
\begin{equation}
\delta_{***} \simeq 1.37\lcurv \ \ \ \text{(III $\rightarrow$ IV)} .
\label{delta***}
\end{equation}
Interestingly, Eq.~(\ref{constant}) for the force in regime IV is the same maximal force that a rigid disc of radius $R$ can support before sinking \cite{Vella15a}---despite the drastically different interfacial geometries they produce, neither is significantly better at staying afloat.

\subsection{Geometric model and localization transition}\label{sec:geom}
Here we show that in regimes III and IV, the profile $\zeta(r)$ can be obtained from purely geometric arguments. 
In regime III, this approach recovers the profile found in Ref.~\citealp{Vella18}; in regime IV, we show that softening is accompanied by localization of $\zeta(r)$.

\begin{figure}[b]
\centering
\includegraphics[width=6.0cm]{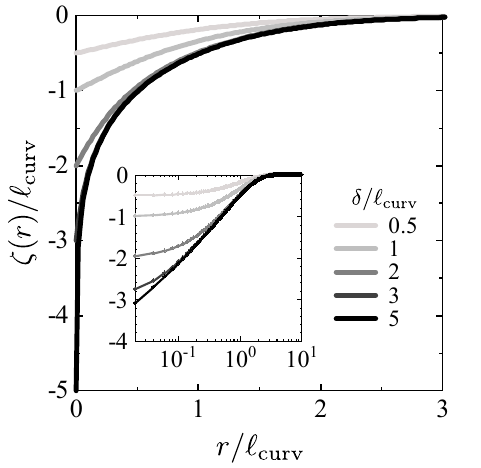}
\caption{
\textbf{Localization at large slopes.} 
Vertical profile of the film, $\zeta(r)$, computed by numerically minimizing Eq.~(\ref{geom2}). 
The profile becomes increasingly localized at large $\delta$. 
\textit{Inset:} Same curves on semilogarithmic axes, showing the approach to a universal curve. 
}
\label{fig:3}
\end{figure}

We treat the sheet as inextensible but with zero bending cost. 
Such a treatment was used to explain the wrapping of a droplet in a thin sheet \cite{Paulsen15} and the folding of an annular sheet submitted to different inner and outer surface tensions \cite{Paulsen17}, and it is motivated by the weak lateral tension and negligible bending stiffness of these films ($\gamma/Y \rightarrow 0$ and $\epsilon \rightarrow 0$). 
In this asymptotic regime, the only relevant energies are due to gravity and surface tension:
\begin{align}
U & = U_\text{gravity} + \gamma (\Delta A_\text{free}) \label{geom}\\
& = \pi\int_0^\infty \left[\rho g r \zeta(r)^2+2\gamma \Rf \left(\sqrt{1+\zeta'(r)^2}-1 \right) \right]dr, \label{geom2}
\end{align}
where $\Delta A_\text{free}$ is the area of the water bath that is exposed by the inward displacement of the sheet and $\zeta(r)$ is the axially-symmetric height profile of the sheet that averages over wrinkles or other microstructures (\textit{i.e.,} the ``gross shape'' of the sheet \cite{Paulsen15}). 
The boundary condition from the indenter is: $\zeta(0)=-\delta$. 
The corresponding Euler-Lagrange equation satisfied by the profile that minimizes the energy is: 
\begin{equation}\label{E-L}
\zeta''(r)=\frac{r\zeta(r)}{\lcurv^3} [1+\zeta'(r)^2]^{3/2} . 
\end{equation}

In the small-slope limit, Eq.~(\ref{E-L}) reduces to $\zeta''(r)=r\zeta(r)/\lcurv^3$, which is solved by an Airy function [see Eq.~(\ref{Airy}) above]. 
For arbitrary slope we solve Eq.~(\ref{E-L}) numerically, yielding the height profiles shown in Fig.~\ref{fig:3}. 
At large indentation, $\zeta(r)$ becomes increasingly localized. 
(One way to see this is observing the deflection at $r/\ell_\text{curv}=1$, which saturates for large $\delta$.) 
Moreover, the profiles appear to approach a universal curve as $\delta$ increases. 
By analyzing the asymptotic behavior of Eq.~(\ref{E-L}), we find that $\zeta(r)\sim -[3\log (1/r)/2]^{2/3}$ as $r\to 0$, in good agreement with the numerical solution (see Fig.~S1). 
As the profile approaches this curve, the volume of fluid lifted by the sheet reaches a plateau, 
so that the gravitational energy eventually becomes negligible compared to surface energy. 
The total energy at large amplitude is thus $U \sim 2\pi \gamma \Rf \delta$, which accounts for the interfacial area of the bath that is exposed as the boundary of the film is pulled inwards. 
This energy scaling recovers Eq.~(\ref{constant}) for the force in regime IV.

\begin{figure*}[b]
\centering
\includegraphics[width = \linewidth]{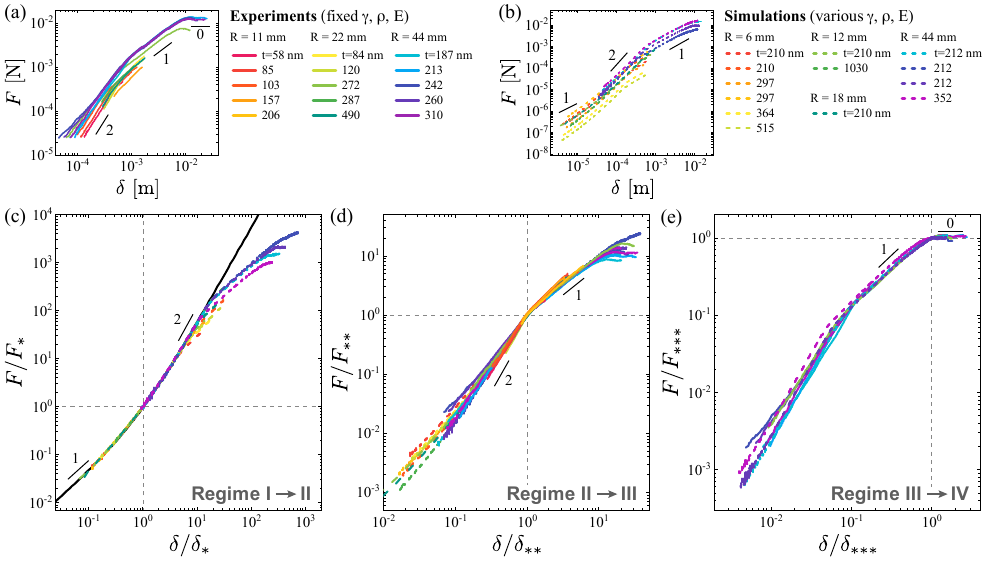}
\caption{
\textbf{Normal force measurements showing four scaling regimes.} 
\textbf{(a,b)} Force versus indentation depth for (a) experiments and (b) simulations at several film radii and a wide range of thicknesses. 
(Values of $\gamma$, $\rho$, and $E$ are listed in the legend to Fig.~\ref{fig:5}.) 
\textbf{(c)} The data at small indentation (simulation curves) are collapsed by rescaling the vertical and horizontal axes by $\delta_{*}$ and $F_{*}$ for each film. 
The data are described well by the empirical form: $F/F_{*} = \frac{1}{2}[\delta/\delta_{*} + (\delta/\delta_{*})^2]$ (black line), until they peel away at large $\delta$. 
\textbf{(d)} A different rescaling by $\delta_{**}$ and $F_{**}$ collapses the data at larger indentation 
in simulations and experiments. 
\textbf{(e)} A transition to a plateau in the force occurs at $\delta_{***}$ in simulations and experiments. 
}
\label{fig:4}
\end{figure*}

\begin{figure*}[b]
\centering
\includegraphics[width=\textwidth]{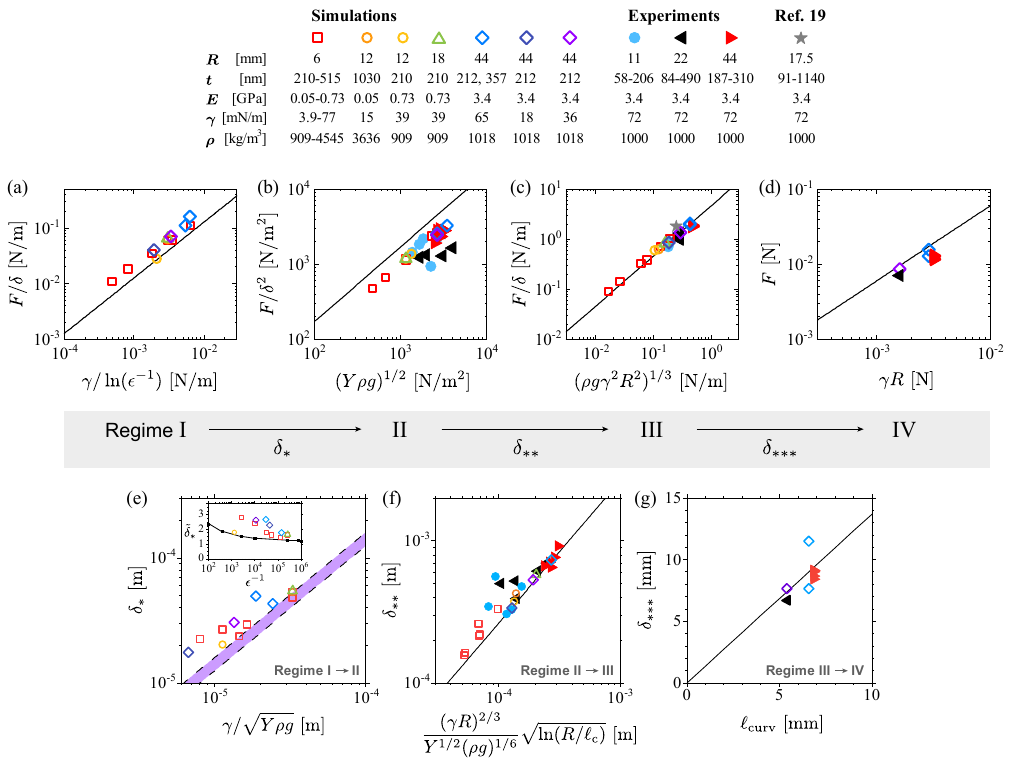}
\caption{
\textbf{Comparison between experiments, simulations, and theory for the magnitude of the force in regimes I-IV (a-d) and the transitions between the regimes (e-g).} 
We varied the sheet and liquid parameters over a wide range, as shown in the legend. 
Panels (a),(e) include only simulations; all other panels include simulations and experiments. 
In all panels, the solid lines show the theoretical predictions given in the main text with no free parameters [Eqs.~\ref{linear}-\ref{delta**} from Refs.~\cite{Vella15,Vella18} and Eqs.~\ref{constant}-\ref{delta***} predicted by our work]. 
The prediction for panel (b) was computed using $\tilde{\delta}=6$, which is in the middle of the range of $\tilde{\delta}$ for the measurements. 
The star in panel (c) shows the value obtained from previous experiments in regime III, reported in Ref.~\citealp{Holmes10}. 
The dashed lines in panel (e) show Eq.~(\ref{delta*}) with the lowest and highest theoretically-predicted prefactors for the simulation parameters used here. 
The inset to panel (e) shows the dimensionless indentation depth $\tilde{\delta}_* = (\delta_*/\ell_\text{c}) \sqrt{(Y/\gamma)}$ versus bendability, $\epsilon^{-1}= \gamma^2 / (B\rho g)$, and the solid line connects points predicted by Ref.~\citealp{Vella18}. 
}
\label{fig:5}
\end{figure*}

\section{Comparison to experiments and simulations}\label{sec:comparison}

We show the measured force curves in experiments and simulations in Figs.~\ref{fig:4}a,b, respectively, where the properties of the film and bath were varied while staying in the highly bendable yet inextensible limit ($\epsilon < 10^{-3}$ and $\gamma/Y < 10^{-3}$). 
We varied the radius and thickness of the film 
over a wide range, as shown in the legend. 
In the simulations, we additionally varied the Young's modulus of the film ($0.05 < E < 3.4$ GPa), as well as the liquid density ($909 < \rho < 4545$ kg/m$^3$) and surface tension ($3.9 < \gamma < 77$ mN/m). 

\subsection{Collapsing the force curves} 
The data over a wide range of parameters may be organized into four distinct scaling regimes, which correspond to regimes I-IV described above. 
At small $\delta$ we can collapse the data by rescaling the axes, $\delta \rightarrow \delta/\delta_{*}$ and $F \rightarrow F/F_{*}$ as shown in Fig.~\ref{fig:4}c, where $\delta_{*}$ and $F_{*}$ are selected for each curve to produce the best collapse. 
Here we rely solely on the simulations to characterize the normal force, as the experiments could not definitively resolve the incipient regime. 
We find an excellent fit to the empirical form: $F/F_{*} = \frac{1}{2}[\delta/\delta_{*} + (\delta/\delta_{*})^2]$, consistent with a broad transition from $F\propto \delta$ to $F\propto \delta^2$ with increasing $\delta$. 
(There is at present no theoretical explanation for this particular form of the crossover between these two regimes.) 

\begin{figure}[b]
\centering
\includegraphics[width=0.35\linewidth]{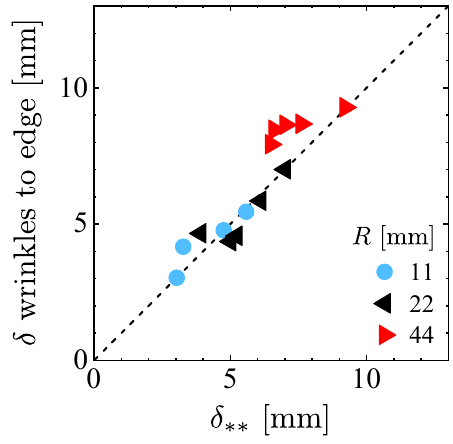}
\caption{
\textbf{Morphological transition marking the start of regime III.}
Indentation depth where wrinkles reach the edge of the film, versus $\delta_{**}$ extracted from the force measurements. 
These two transitions are found to coincide, as predicted in Ref.~\cite{Vella18} and shown by the dashed line with slope 1. 
Data are from experiments using three different film radii. 
}
\label{fig:6}
\end{figure}

At intermediate $\delta$, we observe a transition from $F\propto \delta^2$ to $F\propto \delta$ in the simulations and experiments, which coincides with wrinkles reaching the edge of the sheet [Fig.~\ref{fig:6}]. 
We can collapse all the data in the neighborhood of this transition at $\delta_{**}$ by rescaling the axes, $\delta \rightarrow \delta/\delta_{**}$ and $F \rightarrow F/F_{**}$, as shown in Fig.~\ref{fig:4}d. 
At larger $\delta$, the force reaches a plateau; we collapse the curves around this third transition by selecting $\delta_{***}$ and $F_{***}$ for each curve [Fig.~\ref{fig:4}e].

\subsection{Magnitude of the force} 
This observed sequence of scalings of the normal force with $\delta$ follows the theoretically-predicted progression given by regimes I-IV [Eqs.~(\ref{linear}-\ref{quasilinear}) from Refs.~\citealp{Vella15,Vella18} and Eq.~(\ref{constant}) from the present work]. 
We now show that the data also follow the predicted dependence of the force on the other system parameters, namely, $\Rf$, $t$, $E$, $\gamma$, and $\rho$. 
To reveal this dependence, in Fig.~\ref{fig:5}(a-d) we plot the magnitudes of $F/\delta$, $F/\delta^2$, $F/\delta$, and $F$ measured in regimes I-IV, respectively, which were obtained by measuring the coefficients of these scalings for each sheet. 
(We also include the value measured by Ref.~\citealp{Holmes10} in regime III.) 
We are able to resolve regimes II-IV in experiments, and we find reasonable agreement with the theoretical predictions with no free parameters (solid lines). 
The experimental data in regime II show a systematically lower coefficient than the prediction, whereas the agreement in regime III is excellent. 
We note that in regime IV, the sheet contacts itself over an appreciable area, due to the formation of radial folds. 
Because of this additional adhesion energy that we do not include in the geometric model, we expect the predictions to overestimate the indentation force at large $\delta$, in agreement with observations. 

Our simulations greatly expand the range of the tested parameters, as they allow us to vary $\gamma$ and $\rho$ over a wide range. 
The simulations can also resolve regime I and measure its coefficient [Fig.~\ref{fig:5}a]. 
The data are in reasonable agreement with the prediction, Eq.~(\ref{linear}), but with a somewhat higher numerical prefactor. 
The coefficient in regime II is again lower than the prediction, although the simulation data suggest that the discrepancy is limited to the numerical prefactor in Eq.~(\ref{quadratic}), as the open symbols are shifted down by a constant factor from the solid curve. 
We again find excellent agreement with the prediction in regime III, and good agreement in regime IV.

\subsection{Transitions between the regimes} 
Figure~\ref{fig:5}e shows the indentation depth $\delta_*$ where we observe the transition from regime I to II, obtained by collapsing the simulation data in Fig.~\ref{fig:4}c. 
To compare with theory, we note that Ref.~\cite{Vella18} predicted a numerical prefactor for the scaling in Eq.~(\ref{delta*}) that varies slowly with the bendability, $\epsilon^{-1}$. 
For the physical parameters used here, this prefactor varies from 1.25 to 1.6. 
We show this range with a purple band. 
The data are in reasonable agreement with the prediction. 
To further examine the variation of the numerical prefactor in Eq.~(\ref{delta*}), the inset to Fig.~\ref{fig:5}e shows $\tilde{\delta}_* = (\delta_*/\ell_\text{c}) \sqrt{(Y/\gamma)}$ versus the bendability, $\epsilon^{-1}$. 
The measurements are systematically higher than the prediction obtained from Ref.~\citealp{Vella18}, which has no free parameters. 
Both show a weak dependence on $\epsilon^{-1}$. 

Figure~\ref{fig:5}f shows the measured indentation depth $\delta_{**}$, obtained by collapsing the data in Fig.~\ref{fig:4}d. 
The data are in good agreement with the theoretical prediction, Eq.~(\ref{delta**}), over a wide range of parameters. 
This transition into regime III is predicted to be brought on by the change in the stresses when wrinkles reach the edge of the film~\cite{Vella18}, leading to the Eq.~(\ref{eq:srr_III}) for $\sigma_{rr}(r)$. 
Our experiments provide simultaneous optical, force, and depression measurements, so we can test this scenario directly. 
Figure~\ref{fig:6} compares the indentation depth where wrinkles first reach the edge of the film, with the indentation depth where we see a crossover from a quadratic to a linear scaling in the normal force, \textit{i.e.,} $\delta = \delta_{**}$. 
The data follow the dashed line of slope 1, indicating that these two events indeed coincide. 

Figure~\ref{fig:5}g shows the measured transition depth $\delta_{***}$ marking the onset of regime IV, obtained by collapsing the data in Fig.~\ref{fig:4}e. 
Only a subset of our experiments and simulations were carried out to large amplitude, so we have fewer measurements of this transition. 
Nevertheless, we find good agreement between experiments, simulations, and our theory with no fitting parameters [Eq.~(\ref{delta***})].

\begin{figure}[b]
\centering
\includegraphics[width=0.5\linewidth]{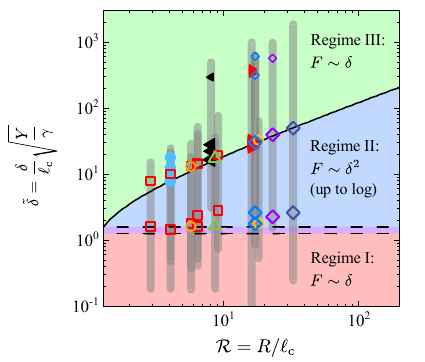}
\caption{
\textbf{Phase diagram for regimes I-III.} 
Dashed lines: Phase boundary for the I $\rightarrow$ II transition, given by Eq.~(\ref{delta*}) for $\delta_{*}$ with numerical prefactors of 1.25 and 1.6, corresponding to the range of $\epsilon^{-1}$ in the data. 
Solid line: Phase boundary for the II $\rightarrow$ III transition, given by Eq.~(\ref{delta**}) for $\delta_{**}$. 
The predictions are supported by our experiments (filled symbols) and simulations (open symbols) over a wide range of parameters (see Fig.~\ref{fig:5} for symbol legend). 
Small symbols mark the observed transition into regime IV. 
}
\label{fig:7}
\end{figure}

Figure~\ref{fig:7} assembles regimes I-III into a phase diagram using the dimensionless parameters $\tilde{\delta} = (\delta/\ell_\text{c}) \sqrt{(Y/\gamma)}$ and $\mathcal{R} = \Rf/\ell_\text{c}$. 
Regime IV is not included in the diagram, because the III $\rightarrow$ IV transition is determined by a different dimensionless group, $\tilde{\tilde{\delta}} = \delta/\lcurv$. 
Nevertheless, our predictions imply that $\delta_{***} > \delta_{**}$ as long as $Y/\gamma \gg (\Rf/\ell_\text{c})^{2/3}$, which is the case in our studies (as well as most experimentally-accessible scenarios in the inextensible regime, $Y/\gamma \gg 1$).

\begin{figure}[b]
\centering
\includegraphics[width=0.6\linewidth]{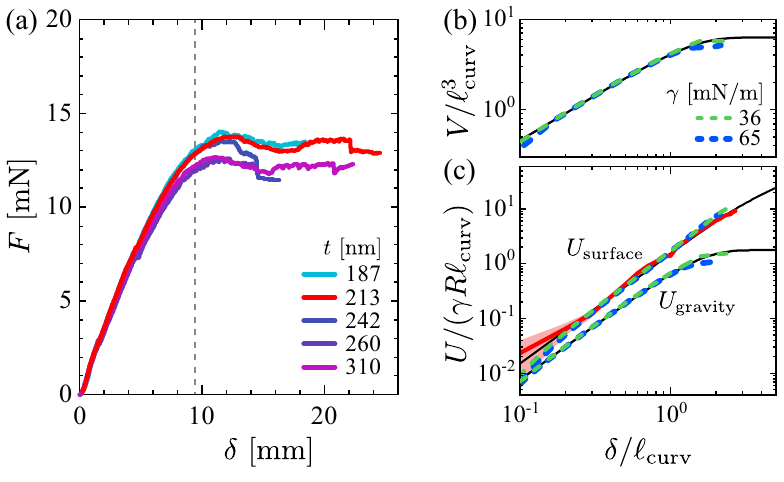}
\caption{
\textbf{Softening and localization at large slopes.} 
\textbf{(a)} Normal force versus displacement in experiments with $\Rf=44$ mm and a range of thicknesses. 
At large $\delta$, the force reaches a plateau that shows no trend with thickness. 
This transition to regime IV is captured by Eq.~(\ref{delta***}) for $\delta_{***}$ (dashed line). 
\textbf{(b)} The volume of the fluid that is lifted by the film, $V$, plateaus to a value proportional to $\lcurv^3$. 
Solid line: Axisymmetric theory. 
Dashed lines: Simulations with $t=212$ nm, $\Rf=44$ mm, and two different values of surface tension. 
\textbf{(c)} Surface energy becomes dominant at large $\delta$ when the lifted volume saturates. 
The axisymmetric predictions for $U_\text{surface}$ and $U_\text{gravity}$ (black curves) are in excellent agreement with our simulations (dashed lines) and experiments (solid red line: $t=213$ nm, $\Rf=44$ mm). 
}
\label{fig:8}
\end{figure}

\subsection{Softening and localization}
To examine regime IV in more detail, Fig.~\ref{fig:8}a shows $F(\delta)$ on linear axes for a set of experiments with $\Rf = 44$ mm. 
At large displacement, the force reaches a plateau value that shows no trend with sheet thickness. 
We mark the predicted transition at $\delta_{***} \simeq 9.4$~mm with a dashed line, which is in excellent agreement with the data. 
The force is predicted to saturate at $\sim$$20$ mN from Eq.~(\ref{constant}); this is the order of magnitude of the observed plateau, but other effects such as folding \cite{Holmes10} should play a role in determining the value of the saturation force, as discussed above. 

Figure~\ref{fig:8}b shows the volume displaced by the film in two simulations carried out to large amplitude with differing values of the surface tension. 
The curves are collapsed by plotting the volume scaled by $\lcurv^3$ and the indentation depth scaled by $\lcurv$. 
The quantitative evolution is described extremely well by our geometric model, including the plateau predicted at large $\delta$, which results from the localization of the vertical profile at large amplitude. 
As the displaced volume saturates, our theory predicts that the surface energy, $U_\text{surface} = \gamma(\Delta A_\text{free})$, becomes dominant over the gravitational energy, $U_\text{gravity}$ defined in Eqs.~(\ref{geom}-\ref{geom2}). 
This progression is supported by Figure~\ref{fig:8}c, where we show $U_\text{surface}$ and $U_\text{gravity}$ measured in the same simulations, in agreement with our theory. 
We also obtain $U_\text{surface}$ in experiment by measuring the shape of the boundary of the sheet in top-view images, further corroborating this picture.

\subsection{Crumples and folds}
Two additional morphologies are observed in the experiments: stress-focusing ``crumples'' that consist of repeated buckled structures terminating at sharp tips \cite{King12,Timounay20}, and radial folds where the film contacts itself \cite{Holmes10}. 
Remarkably, crumples do not affect the force on the probe (see Fig.~S5), whereas a fold elicits a small drop in the force. 
According to the geometric model, crumples should only affect the force on the probe if they alter the gross shape of the film from the optimal axisymmetric shape. 
In contrast, folds may contribute an additional energy due to self-contact, which is beyond the scope of this work.

\begin{figure*}[b]
\centering
\includegraphics[width=12.25cm]{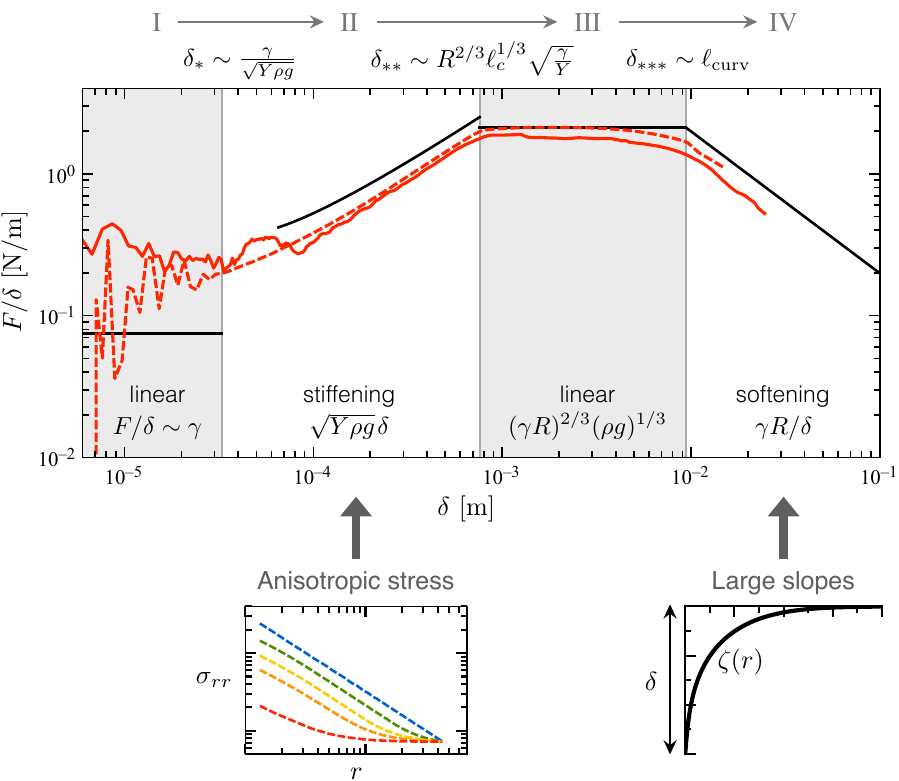}
\caption{
\textbf{Stiffening and softening of a single film.} 
Stiffness, $F/\delta$, versus displacement, measured in experiment (solid red line) and simulation (dashed line) for $\Rf=44$ mm, $t=213$ nm, $E=3.4$ GPa, $\gamma=72$ mN/m, and $\rho=1000$ kg/m$^3$. 
Previous work~\cite{Vella15,Vella18} predicted the stiffness in regimes I-III via the F\"oppl--von K\'arm\'an equations [black lines given by Eqs.~(\ref{linear}), (\ref{quadratic}), and (\ref{quasilinear})]. 
Our geometric model reproduces the result in regime III and predicts a new regime with softening behavior [rightmost black line: Eq. (\ref{constant})]. 
Vertical lines show values of $\delta_{*}$, $\delta_{**}$ predicted by Ref.~\citealp{Vella18} [Eqs.~(\ref{delta*}), (\ref{delta**})] and $\delta_{***}$ predicted by our work [Eq.~(\ref{delta***})]. 
The figure contains no adjustable parameters. 
As a reference, scalings for $F/\delta$ and transitions are written up to logarithmic corrections. 
}
\label{fig:9}
\end{figure*}

\section{Conclusion} 

\noindent Despite the simplicity of the poking protocol, we have shown that it gives rise to two distinct geometric nonlinearities. 
Stiffening is controlled by the dimensionless number: $\tilde{\delta} \equiv (\delta/\ell_c) \sqrt{Y/\gamma}$, which compares geometry-induced strain to the mechanical strain imposed by surface tension, as predicted by Ref.~\citealp{Vella18}. 
Our results at large amplitude have identified a softening response that is controlled by the dimensionless number: $\tilde{\tilde{\delta}} \equiv \delta/\lcurv = \delta/(\ell_c^{2/3} \Rf^{1/3})$, which approximates the maximal slope of the sheet. 
Figure \ref{fig:9} summarizes our results by showing the evolution of the stiffness, $F/\delta$, using simulation, experiment, and theory, for a film with $t=213$ nm. 
The stiffness may be seen transitioning through the four regimes as the indentation depth is increased. 

The mechanism for the observed stiffening is the growing anisotropy of stresses in the sheet as loading progresses. 
The resulting hoop compression eventually leads to wrinkles, which for very thin films cause the compressive stresses in the azimuthal direction to vanish, ($\stt=0$), so that radial forces are passed directly through the wrinkled region, rather than being carried evenly by the radial and hoop components. 
This leads to a long-range decay of the radial stress, $\srr(r) \propto 1/r$. 
The basic physical ingredients leading to this enhanced force propagation are quite modest, so that similar behaviors occur in other materials. 
For instance, the same stress field arises in the mechanics of fiber networks, which make up paper, textiles, and the structural components of tissues and cells \cite{Fletcher10}. 
There, the buckling of fibers sets a threshold for the maximum compressive stresses that may be endured, which is a natural analogue to the formation of wrinkles in thin sheets subjected to small compressional loads. 
This applies to both two- and three-dimensional networks; in $d$ dimensions, Eq.~(\ref{eq:srr_III}) becomes $\srr(r) \propto r^{-(d-1)}$ \cite{Rosakis15}. 
(The stress field for forcing at multiple sites does not obey superposition, so the behavior will differ at high density of forcing sites \cite{Ronceray16}.) 

Softening is not related to the stress field but rather to the gross shape---it arises when there are large deviations from the original slopes on the surface of the object. 
Seen in this way, one may identify similarities with other interfacial problems. 
For instance, when a thin film is laterally compressed on a liquid bath, the confining force softens as it undergoes a wrinkle-to-fold transition with large slopes~\cite{Pocivavsek08,Demery14}. 
The energy functional [Eq.~(\ref{geom2})] is qualitatively similar to that of a \textit{liquid} meniscus in 1D and 2D, where the profile also localizes at large displacements \cite{Anderson06}. 

Recent work has proposed using elastic sheets to tailor the mechanical, chemical, or optical properties of droplets and interfaces \cite{Kumar18}, in analogy with molecular or particulate surfactants \cite{Binks02}. 
Here we have shown how a simple geometric model---originally developed to understand shape selection \cite{Paulsen15,Paulsen17}---can also predict forces, including at large displacements where conventional approaches may fail. 
In particular, Eq.~(\ref{geom}) may be used to determine the energy-minimizing configurations of a sheet as a function of a continuously-varying control parameter (here, the indentation depth $\delta$), which then determines the force. 
This general approach is suitable for other forms of loading such as displacements applied at the edge of a sheet, and it applies to intrinsically curved sheets and curved liquid surfaces.  
This versatile method thus opens the way for understanding the mechanics of sheet-laden interfaces in general settings. 


\section*{Acknowledgements}
We are grateful to Benny Davidovitch and Dominic Vella. 
We thank Douglas Holmes for sharing his force probe design. 
Simulations were performed at the Triton Shared Computing Cluster at the San Diego Supercomputer Center and the Comet cluster (Award no.~TG-MSS170004 to T.Z.) in the Extreme Science and Engineering Discovery Environment. 
Funding support from National Science Foundation Grants No.~IGERT-1068780 (M.M.R.), No.~DMR-CAREER-1654102 (M.M.R. and J.D.P.), and No.~CMMI-CAREER-1847149 (T.Z.) is gratefully acknowledged.

%




\newpage
\clearpage

\begin{adjustwidth}{-0.5cm}{-0.5cm}

\renewcommand{\thefigure}{S\arabic{figure}}
\setcounter{figure}{0} 
\renewcommand{\thetable}{S\arabic{table}}
\setcounter{table}{0} 
\renewcommand{\theequation}{S\arabic{equation}}
\setcounter{equation}{0}

\noindent\textsf{\textbf{\LARGE Supplementary Information for \\}} \\
\noindent\textsf{\textbf{\Large Geometry underlies the mechanical stiffening and softening of an indented floating film \\}} \\
\noindent\textsf{\textbf{\footnotesize Monica M. Ripp, Vincent D\'emery, Teng Zhang, and Joseph D. Paulsen}}
\vspace{0.5in}

{\setstretch{1.0}\footnotesize

\SIsection{Supporting Information Text}\label{}

\SIsection{Geometric approach: Energy functional}\label{}

\noindent The axisymmetric configuration is described by a function $\zeta(r)$, which describes the height of the sheet.
The boundary condition at $r=0$ is given by the poking amplitude,
\begin{equation}
\zeta(0) = -\delta.
\end{equation}
The sheet extends up to a radius $W$, which is given by length conservation:
\begin{equation}\label{eq:def_W}
\int_0^W\sqrt{1+\zeta'(r)^2}dr = R.
\end{equation}
In general, the boundary condition at $W$ is given by continuity relations, and the profile of the liquid surface should be solved for. 
Here, we assume that the length over which the sheet is deformed is much smaller than $R$ and we can write $\zeta(W)=0$.

The energy is the sum of the gravitational energy,
\begin{equation}
U\ind{gravity}=\pi\rho g\int_0^W r \zeta(r)^2 dr,
\end{equation}
and the surface energy, which is given by the excess area of the exposed liquid interface:
\begin{equation}
U\ind{surface}=\pi\gamma(R^2-W^2);
\end{equation}
we choose the convention so that $U\ind{surface}=0$ in the flat state.

Since we have assumed that $\zeta(r)\simeq 0$ for $r\sim W$, we can extend the function $\zeta(r)$ over $[0,\infty]$, and we can write
\begin{align}
U\ind{gravity}&=\pi\rho g \int_0^\infty r \zeta(r)^2 dr, \label{eq:ug}\\
R-W&=\int_0^\infty \left[\sqrt{1+\zeta'(r)^2}-1\right]dr.
\end{align}
If we assume that the inward motion of the edge of the film is much smaller than its radius, we get for the surface energy:
\begin{equation}\label{eq:us}
U\ind{surface}=\pi\gamma(R+W)(R-W)\simeq 2\pi\gamma R\int_0^\infty \left[\sqrt{1+\zeta'(r)^2}-1\right]dr.
\end{equation}
Finally, the total energy is
\begin{equation}
U=U\ind{gravity}+U\ind{surface} = \pi\int_0^\infty \left( \rho gr\zeta(r)^2+2\gamma R \left[\sqrt{1+\zeta'(r)^2}-1\right]\right) dr,
\end{equation}
and the boundary conditions are $\zeta(0)=-\delta$, $\lim_{r\to\infty}\zeta(r)=0$.

\SIsection{Limiting shape at large indentation}\label{}

\noindent Using $\ell\ind{curv}$ as the unit length, the Euler-Lagrange equation for the profile reads
\begin{equation}
\zeta''(r)=r\zeta(r)[1+\zeta'(r)^2]^{3/2}.
\end{equation}
Here we determine the asymptotic behavior of the solution as $r\to 0$ in the limit of infinite confinement, $\lim_{r\to 0}\zeta(r)=-\infty$.
In this limit, $|\zeta'(r)|\gg 1$ and the equation reduces to
\begin{equation}
\zeta''(r)=r\zeta(r)\zeta'(r)^3.
\end{equation}
Writing the profile as $\zeta(r)=f(\log(r))$, the equation for $f(u)$ is
\begin{equation}
f''(u)-f'(u)=f(u)f'(u)^3.
\end{equation}
We have to determine which term dominates in the left hand side. 
If $f''(u)$ dominates, we arrive at $f''=ff'^3$, leading to $f(u)\sim u^{1/3}$; but then $f''(u)\ll f'(u)$ as $u\to\infty$, which is in contradiction with our assumption.
We should thus assume that $f'(u)$ dominates, leading to $f(u)f'(u)^2=-1$, which is solved by $f(u)=-(3u/2)^{2/3}$, and
\begin{equation}\label{eq:univ_shape}
\zeta(r)\underset{r\to\infty}{\sim} - \left[\frac{3}{2}\log \left(\frac{1}{r} \right) \right]^{2/3}.
\end{equation}
This asymptotic shape is in very good agreement with the numerical integration of the Euler-Lagrange equation (Fig.~\ref{fig:univ_shape}).

\SIsection{Simulation method}\label{}

\SIsubsection{Lattice model}\label{}
We developed a lattice based numerical model where the elastic sheet is described by a triangular lattice model \cite{Seung88}, and the liquid surface tension is described by a spring with zero rest length \cite{Giomi12}. 
The total elastic energy of the triangular lattice model can be defined as a combination of the stretching energy and bending energy,
\begin{equation}
U_\text{sheet} = \frac{\sqrt{3}}{4} Y \sum_{ij} (r_{ij}-r_0)^2 + \frac{2}{\sqrt{3}} B \sum_{\alpha\beta}(1-\mathbf{n}_\alpha \cdot \mathbf{n}_\beta )
\end{equation}
where Y is the in-plane stiffness, B is the bending stiffness, $r_{ij}$ is the current bond length, $r_0$ is the equilibrium bond length, and $\mathbf{n}_\alpha$ and $\mathbf{n}_\beta$ are the normal vectors of nearest neighbors (Fig.~\ref{fig:sims_method}a). 
The in-plane stiffness $Y=E t$ and bending stiffness $B = E t^3/(12 (1-\Lambda^2))$ of the thin sheet are defined in terms of the Young's modulus $E$, Poisson's ratio $\Lambda$, and film thickness $t$. 

The liquid surface tension is modeled as zero-rest length spring, which tends to minimize the spring length and thus the total surface area. 
If the springs form an equilateral triangle, the spring constant can be directly linked to the surface tension as:
\begin{equation}
U_\text{liquid} = \frac{1}{2\sqrt{3}} \gamma \sum_{ij} r_{ij}^2.
\end{equation}
We adopted a high resolution lattice model to make sure the deviations of the equilateral triangles are small. 
The gravity force was directly applied to the particles in the elastic sheet, 
\begin{equation}
F_\text{gravity} = -\frac{\sqrt{3}}{2} r_0^2 \rho g z,
\end{equation}
where the force was only along $z$ direction, and the coefficient represents the effective area of a particle in the triangular lattice model. 

We embedded an elastic sheet in a liquid surface within a square simulation domain with periodic boundary conditions. 
All the simulations were carried out with molecular dynamics software LAMMPS. 
A small spherical indenter of radius $110$ $\mu$m was adopted in the simulation via the command ``fix indent'' in LAMMPS. 
The indenter was slowly moved at a constant speed to poke the elastic thin sheet, ensuring a quasi-static process. 
In the simulations, we first followed the metal units in LAMMPS and then scaled the quantities to physical spaces comparable to experiments. 
The key units directly used in the simulations are summarized in Table~\ref{tab:1}. 
We scale the energy unit by $\alpha$ and the length unit by $\beta$ to map the simulations onto a physical structure comparable to the experimental set up. 
The scaled units are shown in the second row of Table~\ref{tab:1}, with two adjustable parameters $\alpha$ and $\beta$. 
In all the simulations, we set $\alpha=10^8$ and $\beta=1.2 \times 10^5$. 
We use the scaled units in the main text. 

During simulations, a Langevin thermostat was adopted to maintain a very low temperature ($0.001$ K in the simulation units). 
We also reduced the indenter speed for a typical simulation and did not observe significant change in the measured forces (Fig.~\ref{fig:sims_rate}), indicating that the speed of the indenter is sufficiently slow. 

\SIsubsection{Treatment of gravity at large slopes}\label{}
For the films with $\Rf = 44$ mm, we carried out our simulations to large amplitude where the sheet attains large slopes. 
This situation requires a modified treatment of the gravity force. 
A small triangular element in the elastic sheet will be tilted in the current deformed configuration (Fig.~\ref{fig:nonlinear_gravity}), effectively reducing the volume of liquid lifted by the solid film. 
The volume change due to the film deformation can be expressed as:
\begin{equation}
V = A_{xy} \overline{z},
\end{equation}
where $A_{xy}$ is the projected area of the triangle on xy plane, $\overline{z} =1/3(z_1+z_2+z_3)$ is the height of the centroid (O) of the triangle. 
Therefore the gravity energy at large deformation can be written as:
\begin{equation}
U_\text{gravity} = \frac{1}{2}\rho g A_{xy} \overline{z}^2.
\end{equation}

The force components applied on each node in the triangle can be calculated as:
\begin{subequations}
\begin{equation}
f^i_x = - \frac{\partial{U_\text{gravity}}}{\partial{x_i}} = - \frac{1}{2} \rho g \overline{z}^2 \frac{\partial{A_{xy}}}{\partial{x_i}}
\end{equation}
\begin{equation}
f^i_y = - \frac{\partial{U_\text{gravity}}}{\partial{y_i}} = - \frac{1}{2} \rho g \overline{z}^2 \frac{\partial{A_{xy}}}{\partial{y_i}}
\end{equation}
\begin{equation}
f^i_z = - \frac{\partial{U_\text{gravity}}}{\partial{z_i}} = - \frac{1}{3} \rho g \overline{z} A_{xy}
\end{equation}
\end{subequations}
To avoid material penetration, a purely repulsive force is applied to particles in the solid film, such as:
\begin{equation}
f_{ij} = K_c \sin\left(\frac{\pi r}{r_c}\right); \ \ \ (r < r_c),
\end{equation}
where $K_c$ controls the strength of the repulsive force, $r_c$ is the cutoff of the interaction and the force is 0 for $r>r_c$.

\SIsection{Experimental methods and analysis}\label{}

\SIsubsection{Film preparation}\label{}
We made polymer films by spin-coating solutions of polystyrene ($M_\text{n} = 99$k, $M_\text{w} = 105.5$k, Polymer Source) in toluene ($99.9\%$, Fisher Scientific) onto glass substrates, following Ref.~\cite{Huang10}. 
After the indentation, each film (or a portion of the film) was retrieved on a silicon wafer. 
Film thickness was then measured using a white-light interferometer (Filmetrics F3). 
Thicknesses were found to be uniform to within 2\% when $\Rf=11$ or $22$ mm, and to within $3\%$ when $\Rf=44$ mm. 

\SIsubsection{Force measurements}\label{}
We measured normal forces using a custom setup that uses a capacitive sensor (PI PiSeca E-852 with D-510.020) to detect the deflections of a metal cantilever that pushes down on the sheet via a spherical indenter tip. 
This force probe is mounted on a computer-controlled vertical translation stage with a resolution of 5 $\mu$m. 
The apparatus was calibrated by hanging known masses from the indenter. 
We tested this calibration method with an independent measurement where we hung a Wilhelmy plate from the intender tip and lowered it into water (ensuring full wetting of the water to the plate), and we recovered the surface tension of a clean air-water interface to within $1$ mN/m. 

To identify the moment of contact, we first examine the corresponding video to find its approximate time. The precise moment of contact is marked by a significant reduction in noise in the capacitive sensor signal. We set $\delta=0$ to coincide with this noise drop, and $F=0$ is found by averaging the force at earlier times. 

\SIsubsection{Measuring $\delta^*$}
Reference~\cite{Vella18} defined $\delta_*$ as the point where $F/\delta^{3/2}$ is minimized. 
Here we measure $\delta_{*}$ by collapsing the data to the empirical form $F/F_{*} = \frac{1}{2}[\delta/\delta_{*} + (\delta/\delta_{*})^2]$, where $F_{*}$ and $\delta_{*}$ are free parameters for each measured curve. 
So long as there is sufficient data on each side of the crossover, the two methods are essentially equivalent, since the function $(x + x^2) / x^{3/2}$ is minimized at $x=1$. 
However, when determining $\delta_{*}$ by our method, one must ensure that there is sufficient data on each side of the crossover in order to obtain reliable results.

\SIsection{Crumpling transition}

\noindent At large indentation depth beyond $\delta_{**}$, the previously-smooth wrinkled pattern becomes concentrated into a discrete set of deformations, as shown in Fig.~\ref{fig:crumple}a. 
This progression is similar to what is observed when a circular polymer sheet is placed on a droplet of gradually increasing curvature~\cite{Paulsen15}. 
These structures have been termed ``crumples'', and their appearance marks a symmetry-breaking transition that is traversed as the indentation depth is varied, but their underlying physical mechanism is not understood~\cite{King12}. 
Crumples are observed at systematically smaller indentation for thicker films. 

One might expect a signature in the normal force when crumples appear, since they are known to focus stress at their tips~\cite{King13}. 
Surprisingly, the data are featureless through this transition despite the two distinct morphologies, as shown in Fig.~\ref{fig:crumple}b. 
This observation can be partially justified in the far-from-threshold framework \cite{Davidovitch11} by noting that both crumples and wrinkles allow compression with vanishing elastic cost.
 
}

\newpage
\clearpage

\begin{figure}
\begin{center}
\includegraphics[width=9.0cm]{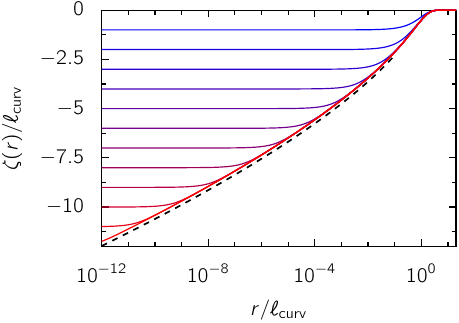}
\end{center}
\caption{Height profile from the numerical solution of the Euler-Lagrange equation for $\delta\in\{1,2,\dots,12\}$ (blue to red solid lines) and asymptotic solution (Eq.~(\ref{eq:univ_shape}), dashed black line).}
\label{fig:univ_shape}
\end{figure}

\begin{figure}
\centering
\includegraphics[width=8.0cm]{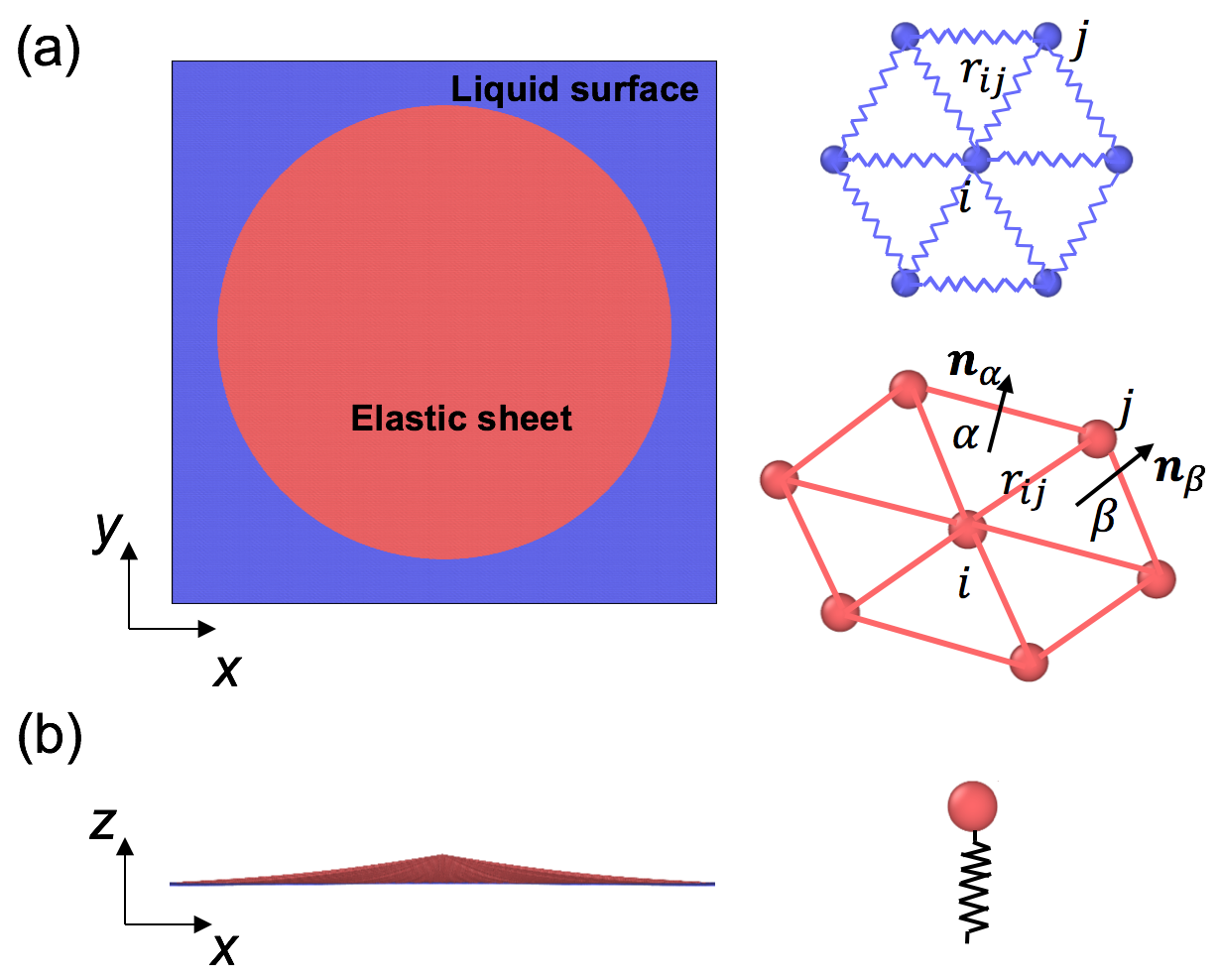}
\caption{
Simulation model. 
\textbf{(a)} Top view of the simulation domain, and lattice elements for the liquid surface and the sheet. 
\textbf{(b)} Side view. Vertical displacements of the particles in the sheet are coupled to linear springs that impose a gravitational force. 
}
\label{fig:sims_method}
\end{figure}

\begin{figure}
\centering
\includegraphics[width=9.0cm]{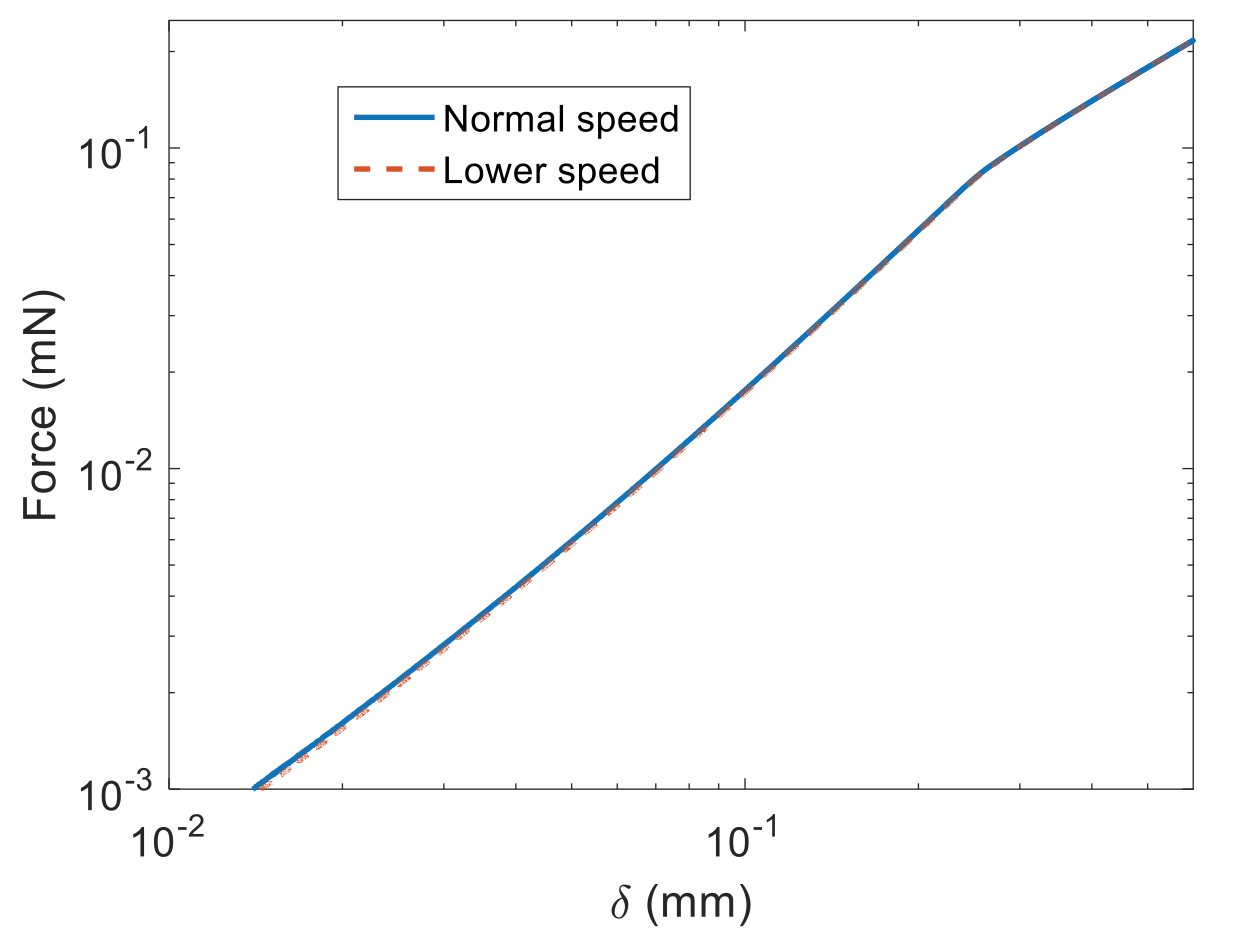}
\caption{
Force versus indentation depth for two indenter speeds, measured in simulations with $\Rf=6$ mm, $t=210$ nm, $E=0.73$ GPa, $\gamma=39$ mN/m, $\rho=909$ kg/m$^3$. 
The curves are in good agreement, consistent with a quasistatic indentation process. 
}
\label{fig:sims_rate}
\end{figure}

\begin{figure}
\centering
\includegraphics[width=7.5cm]{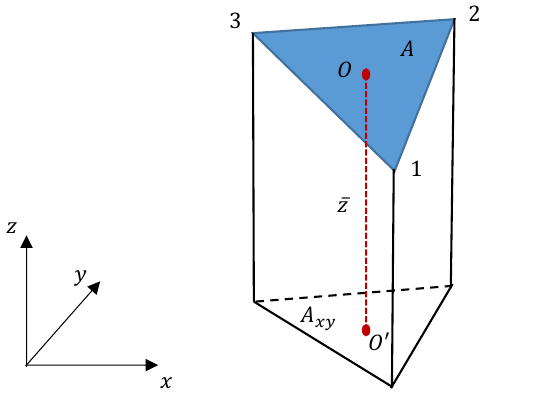}
\caption{
Treatment of gravity in simulations reaching large slopes. 
The solid triangle represents a small element in the elastic thin sheet. 
$O$ is the centroid of triangle, and $O'$ is the centroid of the projected triangle on the $x$-$y$ plane. 
}
\label{fig:nonlinear_gravity}
\end{figure}

\newpage
\clearpage

\begin{figure}
\centering
\includegraphics[width=10.0cm]{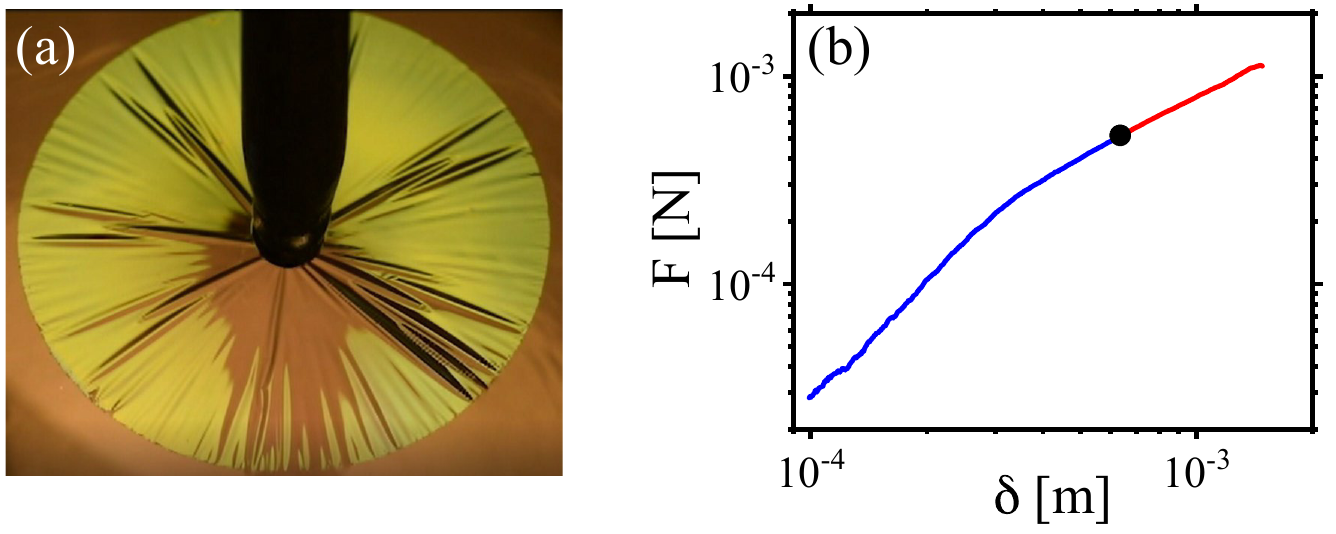}
\caption{
Wrinkle-to-crumple transition. 
\textbf{(a)} Stress-focusing crumples appear at large indentation, as shown here at $\delta=1.08$ mm for a film with $t = 437$ nm and $\Rf = 11$ mm. 
\textbf{(b)} The transition is not apparent in the force. 
(Black circle: $\delta$ where crumples appear.)
Here, $t = 103$ nm and $\Rf = 11$ mm. 
}
\label{fig:crumple}
\end{figure}

\begin{table}
\centering
\caption{
Simulation units in LAMMPS (native) and scaled results.
}
\begin{tabular}{ |c||c|c|c|c|c| }
 \hline
 & Bending stiffness & Length & Surface tension & Young's modulus & Gravity \\
 \hline
 LAMMPS units & eV & \AA & eV/\AA$^2$ & eV/\AA$^3$ & eV/\AA$^4$ \\
 Scaled units & $\alpha$ eV & $\beta$ \AA & $(\alpha/\beta^2)$ eV/\AA$^2$ & $(\alpha/\beta^3)$ eV/\AA$^3$ & $(\alpha/\beta^4)$ eV/\AA$^4$ \\
 \hline
\end{tabular}
\label{tab:1}
\end{table}

\end{adjustwidth}

\end{document}